\documentclass[pre,twocolumn,showpacs,preprintnumbers,amsmath,amssymb]{revtex4}

\usepackage{graphicx}
\usepackage{dcolumn}
\usepackage{bm}

\begin{document}

\preprint{}

\title{Multiple Scale-Free Structures in Complex Ad-Hoc Networks}%

\author{Nima Sarshar}

\author{Vwani Roychowdhury}%
 \email{{nima,vwani}@ee.ucla.edu}
\affiliation{Department of Electrical Engineering, University of California, Los Angeles}%

\date{\today}

\begin{abstract} \small
This paper develops a framework for analyzing and designing
dynamic networks comprising different classes of nodes that
coexist and interact in one shared environment.  We consider {\em
ad hoc} (i.e., nodes can leave the network unannounced, and no
node has any global knowledge about the class identities of other
nodes) {\em preferentially grown networks}, where different
classes of nodes are characterized by different sets of local
parameters used in the stochastic dynamics that all nodes in the
network execute. We show that multiple
scale-free structures, one within each class of nodes, and with
tunable power-law exponents (as determined by the sets of
parameters characterizing each class) emerge naturally in our
model. Moreover, the coexistence of the scale-free structures of
the different classes of nodes can be captured by succinct phase
diagrams, which show a rich set of structures, including stable
regions where different classes coexist in heavy-tailed (i.e.,
exponent is between 2 and 3) and light-tailed (i.e., exponent is
$> 3$) states, and sharp phase transitions. The topology of the
emergent networks is also shown to display a complex structure,
akin to the distribution of different components of an alloyed
material; e.g., nodes with a light-tailed scale-free structure get
embedded to the outside of the network, and have most of its edges
connected to nodes belonging to the class with a heavy-tailed
distribution. Finally, we show how the dynamics formulated in this
paper will serve as an essential part of {\em ad-hoc networking
protocols}, which can lead to the  formation of robust and
efficiently searchable networks (including, the well-known
Peer-To-Peer (P2P) networks) even under very dynamic conditions.
\end{abstract}

\maketitle

\section{\label{sec:level1}Introduction} \vspace*{-6pt}
\subsection{Motivation}\vspace*{-6pt}
Real networks are rarely homogeneous, and often comprise different categories of constituent nodes, all of which interact and coexist in a single global environment. The classification of nodes in such networks might be based on different characteristics, including diverse functionalities, different objectives and scale of resources, and different types and degrees of
dynamics inherent to the nodes. Examples include, different cell types in a neural
network, different roles in the web of English words (e.g., verbs,
nouns, and adjectives), different disciplines in the network of
scientific collaborations, various switch types (routers,
hubs, etc.) in the Internet, and the different node types in a Peer-to-Peer
(P2P) network (e.g., nodes with 56K-Baud modem connections vs. nodes with DSL connections). Empirical evidence of class-specific hierarchical and scale-free structures in many complex networks have become available only recently \cite{hir}.  For example, a heterogeneous scenario has been reported in the network of scientific
citations: While the overall structure of such networks is
believed to have a power-law (PL) degree distribution\footnote{Recall that scale-free structures are often captured by a power-law (PL) degree distribution, where the probability of a randomly chosen
node to have degree $k$ scales as $P(k)\propto k^{-\gamma}$ for
large $k$; $\gamma$ is referred to as the exponent of the distribution.} with exponent $\gamma \approx 3$, the
network of citations restricted to theoretical physicists is somewhat more heavy tailed with an exponent around $\gamma\sim 2.6$
\cite{the}.

The objective of this paper is to explore the {\em dynamics of such heterogeneous networks}, and study how different types of nodes influence each other, and {\em when and how  multiple scale-free structures may  emerge} in the networks. We find that the interacting sets of different classes of nodes can give rise to a complex global structure, and display a rich set of emergent properties. In particular, we show (i) Viewing complex systems as
networks with different evolving, and interacting constituent
subnetworks, helps gain better understanding about the role
of each class of nodes in the overall network, and sheds light on
how different networks with nested structures {\em might have evolved}, and (ii) How the  results can lead to the {\em systematic design of heterogeneous networks}, where  different categories of {\em nodes} evolve to have {\em different scale-free} structures (corresponding to their capabilities and intentions).

\vspace*{-6pt}
\subsection{Dynamical Model and Results}\vspace*{-6pt}
We consider {\em ad hoc} dynamical networks, where in addition to
nodes joining the network, nodes also {\em disconnect and leave
the network randomly} at  certain rates. Moreover, nodes are
allowed to respond to their environment, and can initiate new
random connections if their existing connections are lost. The
dynamical rules studied in this paper, have been picked from those
traditionally studied in the context of complex networks. For
example, in all the protocols studied in this paper {\em all
connections are {\em initiated} by choosing nodes in a linear
preferential manner}. Similarly, the other dynamics that we
incorporate are joining of new nodes, random deletion of existing
nodes, and compensatory rewiring by existing nodes that might have
lost an edge because of the deletion dynamics. As described next,
the different classes of nodes in the network follow the same
global dynamics, except with {\em different parameters}.
Additionally, to be true to the ad hoc nature of our model, we
enforce that the {\em class identity of a node is not a global
knowledge}; instead, a node's identity is expressed only {\em
through its own} dynamics and actions, i.e., in  how it responds
to and initiates contacts with other nodes in the network. Thus, a
node on joining the network makes requests for connection without
any global knowledge about the properties of other nodes in the
network, and the connection requests are made following the
traditional preferential attachment rule.

\vspace*{1ex}

\noindent
{\em Modeling Heterogeneity of Nodes:}   The different classes of nodes are characterized by different sets of parameters they adopt in responding to connection requests, and also in executing local dynamics. The most important {\em local} parameters used in this paper to capture heterogeneity of nodes, include (i) \underline{\em attraction or attachment}, i.e., a node's willingness to accept a requested connection; this can be parametrized  by the probability, $d_q$, with which the nodes in the $q^{th}$ class accept a connection request. (ii) \underline{\em stability}, i.e., how long they stay in the network
before dropping out or deleting themselves; this can be parametrized  by the probability, $c_q$, with which a randomly picked node in the $q^{th}$ class gets deleted at each time step. (iii) \underline{\em responsiveness}, i.e., a node's
ability to respond to lost or dead connections and compensate for them with
new connection; this can be parametrized  by the probability, $n_q$ with which a node in the $q^{th}$ class compensates for any lost or inactive connection, and (iv) \underline{\em representation} or {\em relative population}, i.e., what percentage of the nodes joining the network belong to a specific class;  this can be parameterized  by the probability, $s_q$, with which a node joining the network is from the $q^{th}$ class.
In general, the different categories of nodes in a network may differ
in all four of these parameters.

\vspace*{-6pt}
\subsection{Approach and A Preview of Results}\vspace*{-6pt}
We are interested in finding $P_q(k)$, the degree distributions
within each of the subclasses. That is, $P_q(k)$ is the probability of a
randomly chosen node of type $q$ to have degree $k$; note that all
edges, both intra- and inter-community edges, contribute to the
degree of a node. In
particular, we will show the emergence of scale-free degree
distributions within each class, that is $P_q(k)\propto
k^{-\gamma_q}$.  In general, for a given set of $Q$ classes, the
PL exponent $\gamma_q$ of the $q^{th}$ class is a function of
all the four sets of  parameters. That is, in general we have
$$\gamma_q=f(D,C,N,S) \hbox{\hspace*{4mm} for all
$q=1,\ldots,Q$},$$
where $D=\{d_1,\cdots,d_Q\}$, $C=\{c_1,\cdots,c_Q\}$, $N=\{n_1,\cdots,n_Q\}$, and $S=\{s_1,\cdots,s_Q\}$, are the sets of,
attachment, stability, responsiveness, and representation parameters for the $Q$ different classes of nodes.
We apply the continuous-time rate equation approach to study the dynamics
and derive $\gamma_q$'s.  The coexistence of different classes of networks
will impose certain constraints on the set of $\gamma_q$'s
emerging in the subclasses, and hence, only certain sets of $\gamma_q$'s for
a given choice of the dynamical parameters are feasible.

As discussed in Section~\ref{s3}, it is always possible to compute the $\gamma_q$'s numerically using the rate-equation approach. {\em An exact closed form computation of the various PL exponents} (as a function of the different sets of parameters), however, is difficult to obtain when the classes are {\em heterogeneous with respect to all} the local parameters.  Hence, to develop intuition and to better understand the heterogeneous systems, we study special cases where it is possible to obtain exact formulas for the computation of $\gamma_q$'s. For example, in Section~\ref{s2} we study the case, where the sets are heterogeneous over only the sets $D$ and $S$, i.e., the $q^{th}$ class has attachment rate $d_q$ and relative population $s_q$. Moreover, we allow deletions of random nodes at an overall rate of $c$ (i.e., the deletion rates of the $q^{th}$ class is given as $c_q=c*s_q,$ and cannot be set independently), and no compensation, i.e., $n_1=n_2=\cdots=n_Q=0$.  Eqs. \ref{eq10} and \ref{eq2} provide the exact formula for $\gamma_q$'s for this particular model.  Interesting features include:
 (i) {\em Coupled PL exponents:} The average PL degree is constrained to be greater than $3$.  Thus for example, for $Q=2$, one class can have PL exponent $<3$, while the other one has to have exponent $> 3$. Recall that if $Q=1$ and $c=0$, it is exactly the case of linear preferential attachment, and the exponent is exactly equal to $3$. Hence, by having multiple classes, one can have classes that have {\em heavy tailed} PL degree distributions (i.e., exponent $<3$), even with the linear preferential attachment kernel. In general, for different $c$'s the possible  PL exponents $(\gamma_1,\gamma_2)$ are plotted in Fig.~\ref{fig-gamma12}. (ii) {\em Role of the Deletion Rate $c$:} If $Q=1$ then it is shown in \cite{us} that for any deletion rate $c>0$, the PL exponent is $>3$, and that the exponent increases rapidly with an increase in $c$. As shown in Fig. \ref{fig-gamma12}, for two classes it is possible to have one class with PL exponent $<3$. However, there is always a deletion rate for which both the exponents become $> 3$.  Thus, the {\em heterogeneity of the network} can lead to rich structures, which otherwise {\em do not exist in homogeneous networks.}

The more general case, where we vary three sets of parameters,
$C,N,S$, while the attachment rates are considered to be unity for
all the classes, i.e., $d_1=\cdots=d_Q=1$, is considered in
Section~\ref{s3}. It is best to describe the results in terms of a
{\em phase space}, where the state of a class can be attributed as
described in the following.  Another means of studying the
networks is to look at the embedding of the different classes of
nodes in the overall structure. Both of these macroscopic
approaches and related results are summarized in the following.
\pagebreak

\noindent{\em Emergence and Coexistence of Phases:} The degree
distribution of a particular category of node could be classified
into the following {\em phases} or states, depending on the
exponent, $\gamma$, of its PL distribution: (i) {\em Heavy
Tailed}: $2\leq \gamma< 3$; for such a distribution, the variance
becomes unbounded while the mean remains bounded. It is in this
regime that the corresponding network shows a number of
advantageous properties, such as almost-constant diameter,
efficiently searchable, and resistance to random deletions of
nodes and edges. (ii) {\em Light Tailed:} $\gamma >3$, (iii) {\em
Unstable:} $0<\gamma <2$, this is when the average degree becomes
unbounded, and (iv) {\em Extinction:} when the average degree goes
to zero, i.e., the nodes belonging to this class get disconnected
from the rest. Clearly, one can make a finer division of the range
of the PL exponent and define a larger number of phases.

We investigate issues related to how the different categories of nodes can be in different phases as a function of the parameters of the dynamics.  Also, how will a phase transition within one
subclass (e.g., from a heavy to a light tailed phase) affect the phase of the other classes? As shown in Section~\ref{sec-phases}, we find that the rich set of
solutions of the dynamical model introduced in this paper, can be
captured by {\em succinct phase diagrams}, which show the
coexistence of different phases of different categories of nodes, as a function of the parameters. The results show that the {\em phase space shows all the hallmarks of a rich heterogeneous system}, including
\begin{itemize}
\item Stable regions where different categories of nodes can be in different desired phases (e.g., certain classes in the {\em heavy-tailed} phase, while certain others in the {\em light tailed} phase). These regions have sufficient volume/area so that the resulting degree-distributions are fairly insensitive to the exact choice of the different parameters; see simulation results in Section~\ref{sec-phases}.

\item Regions in the phase space, corresponding to exhaustive combinations of phases that the different classes of nodes can exist, emerge quite naturally.  For example, Figs.~\ref{fig-phases_new} and \ref{fig-phases_2} show that all four possible combinations of light and heavy tailed phases of two categories of nodes are possible.

\item The parameter space has boundaries showing sharp phase transitions. This can allow abrupt transformations and manipulations of  underlying network topology, by changing the dynamical parameters only marginally.

\end{itemize}

\vspace*{2ex}

\noindent
{\em Topology of the Heterogeneous Networks:} We address issues related to how the different categories of nodes get embedded in the network. For example,  one could ask where in the network do nodes of different classes migrate to?  When a particular class of nodes is in the light-tailed phase, then are the corresponding nodes on the outer edge of the network, in the sense
that, most of its connections are to the nodes outside its own class,
or is it in the core of the network? Similarly, almost all complex networks happen to have small diameters, meaning that there is a
short path from any node to any other node. How many of those
paths pass through a given class of nodes?

We find that instead of nodes segregating into clusters of their
own, they get embedded in the network in such a fashion so {\em
that nodes belonging to classes with light-tailed degree
distributions} are connected via a {\em core comprising of nodes
belonging to the heavy-tailed classes}. This gives rise to global
hierarchical networks, where the nodes can choose its position and
functionality by controlling a set of well-defined parameters. For
example, we define a parameter {\em called the capacity}, which is
the ratio of all edges with both end points in  a particular
class, and the total number of edges with any of its end points in
the same class. Then as shown in Fig. \ref{fig:cap} one can vary
the relative capacities of the different classes by varying the
different parameters.  In general, the results show that when a
class has high exponent, then it's capacity is low and as the
exponent decreases the capacity increases.

\vspace*{-6pt}
\subsection{Implications: Discovering and Modeling Complex Dynamics} 
\begin{center}
\begin{figure*}
\includegraphics[width=3.5in,height=2.5in]{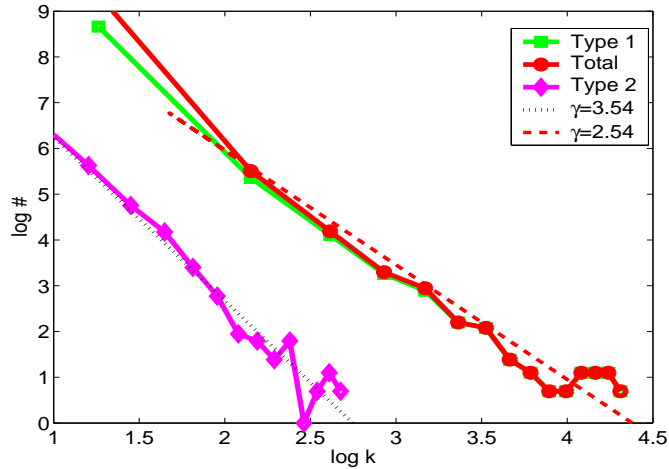}
\caption{\small {\bf An example of a grown heterogeneous network with multiple scale-free structures:}   The network comprises two
classes of nodes, and nodes in both classes follow the same overall preferential dynamics:
when joining the network, every node selects target nodes preferentially until
 a total of $m$ preferential connections are made. The only deviation from the well-known preferential dynamics is that  if a preferentially chosen node
refuses a connection request, then  the requesting node makes
another independent preferential selection and repeats the process
until a request is accepted.  The dynamical rule differentiating
the two classes is that while Type-1 nodes always accept all the
connection requests ($d_1=1$), Type-2 nodes randomly accept only
half the requests  ($d_2=0.5$). The two classes have equal
representation, i.e., $s_1=s_2=0.5$. As shown in the degree
distribution plots, this dynamic leads to two distinct scaling
features: A heavy-tailed degree distribution, $\gamma_1\sim 2.54$,
for Type-1 (squares), and a light-tailed distribution,  $\gamma_2
\sim 3.54$, for Type-2. Interestingly, the overall degree
distribution (circles) closely resembles the degree distribution
of Type-1 nodes, and hence, if the nodes are not differentiated
then the whole network will be characterized by a single scaling
parameter, $\gamma\approx 2.54$. In general, single scaling
parameter characterizations of heterogeneous networks will always
hide the structure of all the classes of nodes, except that of the
most heavy-tailed one: Consider the case of a network consisting
of two classes of nodes, $A$ and $B$, with PL exponents $\gamma_A$
and $\gamma_B$, respectively. If $\gamma_A<\gamma_B$ then the
overall degree distribution is thus $P(k)\sim
k^{-\gamma_A}+k^{-\gamma_B} \sim k^{-\gamma_A}$ when $k \gg
1$.}\label{fig:mixed}
\end{figure*}
\end{center}

\vspace*{-6pt}
An example of how the study of heterogeneous networks may
influence our understanding of mechanisms underlying a given network
is given in
Fig.~\ref{fig:mixed}. It is well known that in a growing network
the standard linear preferential attachment dynamic, where nodes
joining a network makes connections to existing nodes with
probability proportional to their degrees, leads to a degree
distribution with an exponent $\gamma =3$, which marks the
boundary between a heavy-tailed and light-tailed distribution. The
striking characteristics of heavy-tailed degree distributions are
observed only for exponents $< 3$, and most documented networks
display these scalings. In order to account for such widespread
emergence of prominent scale-free structures, several alternate
network dynamics and protocols (e.g., doubly preferential
attachment) \cite{d1,d2,d3,d4,d5,BB}, which lead to a continuum of
possible power-law exponents from $\gamma=2$ to $\infty$, have
been introduced.

Does the presence of an exponent $\gamma <3$
mean that one of the alternate mechanisms is at work? As
illustrated in Fig. \ref{fig:mixed}, {\em the presence of a
heavy-tailed degree distribution need not necessarily imply that
the simple linear preferential attachment is not at play}. Indeed,
a PL degree distribution with exponent less than $3$ can result
from the standard linear preferential attachment dynamic if the
network, for example, has two classes of nodes with varying
acceptance or attachment rates. In this example, a node joining a
network {\em still makes globally preferential links}, except that
when the request reaches one of the classes, it rejects it with
probability $1/2$ while the other class always accepts a
connection for request. From a mechanism perspective, if one did
not view this as a heterogeneous network, then {\em one might be
misled} to infer that the underlying dynamic  was something other
than the standard preferential attachment.

\vspace*{-6pt}
\subsection{Implications: Designer Complex Networks}\vspace*{-6pt}
In terms of  explicitly designing heterogeneous dynamic networks,
an example of great practical interest  is the class of ad-hoc
distributed systems, with peer-to-peer (P2P) content-sharing
networks as a prime example. As discussed in more detail in
Section~\ref{sec-conc}, a P2P network has heterogeneous sets of
nodes with varying lifetimes and bandwidth capabilities. A natural
question is {\em how to design local dynamics} such that an
overall scale-free structure will emerge, where {\em each node
category has a distribution that suits its available resources and
needs}.  We add an additional stringent design constraint: A node
joining the network has no global knowledge of which nodes belong
to which category, and it can only explore the network locally and
only request connections to nodes that it can reach. The dynamics
introduced in this paper {\em provide a systematic solution to
this challenging problem.}

The primary motivations for designing local dynamics so that a  PL
topology emerges include, (i) PL networks are resistant to random
deletions and have vanishingly small percolation thresholds
\cite{zerothresh}; (ii) PL networks have a natural hierarchy
allowing more capable processors to act as hubs; moreover,
computing resources are heterogeneous to begin with and PL
networks provide a natural set-up for the resource hierarchy to be
embedded into a networking hierarchy, and (iii) the structure of
PL networks can be exploited to provide {\em scalable} key-words
based search capabilities \cite{perc,p2p,tcs}. While these
properties of PL networks, have been proven to be true for random
PL networks, our recent results show that the grown random
networks generated using the local dynamics formulated in this
paper (particularly, the ad hoc dynamics, where nodes randomly
leave the network) lead to networks that are {\em much closer to
random PL graphs} than those generated by previously-proposed
algorithms \cite{Callaway01} \cite{unpublished-us}.

\vspace*{-6pt}
\subsection{Prior Work} \vspace*{-6pt}
Previously known dynamical models have mainly characterized the
scale-free structure of the emergent networks with a single state
(manifested by the overall power-law exponent). Nonuniform
preference kernels and their effect on the overall power-law
exponent of the emerging network has bee considered before in the
context of fitness models. In \cite{fitness}, a node dependent
preferential attachment kernel is introduced. As a new node $i$
enters the network, it is assigned a fitness factor $\eta_i$
randomly drawn from some distribution. The probability of the node
$i$ receiving a new connection when its current degree is $k_i$
will be proportional to $\eta_i k_i$.

 The argument is that
different nodes in the network can have different inherent
attractions.  As such, a new node with a high fitness can gain
more connections over time compared to an old node with a smaller
fitness. This can explain, for instance, the high connectivity of
some new pages in the WWW. The authors then derive the {\em overall
degree distribution} as a function of the distribution of the
$\eta_i$'s. The same multiplicative fitness mode is adopted in
\cite{dog-fitness} for the case of directed preferentially grown
graphs. In particular, it is shown that even a single node with a
high fitness can acquire almost all the links in the network over
time, corresponding to a condensation to a star-like topology. The
work  of \cite{dog-fitness} also considers the case of the mixture
of two "weak" and "strong" classes which closely relates to the
model considered in this paper when $Q=2$ and $c=0$. However, {\em a detailed study
of the emergence of different scale-free structures} in heterogeneous networks,  and their dependencies on dynamical parameters {\em have not been addressed prior to this work.}

\subsection{Organization of the paper}
The rest of the paper is organized as follows. In Section
\ref{s1},  we describe a dynamical model of the networks
considered in this paper. We formally introduce the four
parameters that can be used to characterize the different
categories of nodes. In general, all four of these parameters
could be non-uniform over the classes of nodes in the network.
However, for the sake of analysis and also understanding the roles
of different dynamics in determining emergence of scale-free
structures, we study cases where only one or two of these
parameters are nonuniform over the nodes in the network, and the
others are held uniform.  For example, in Section \ref{s2} we
first analyze the model by considering the effects of only
attractions and representations, as these parameters are varied
for different classes of nodes. Moreover, we consider a dynamic ad
hoc network, where nodes both join and leave the network. Next in
Section~\ref{s3}, we solve the model for the case of uniform
attractions,  but heterogeneous stability, responsiveness, and
representation properties. We are interested in explicitly
tracking the structure of each subclass and investigating the
fundamental constraints that the intra-class interactions of a
particular class will impose on its emerging structure. To our
particular interest is the coexistence of different phases in
different classes. In Section~\ref{sec-phases} we explicitly
address this issue and look at the structure of the {\em phase
space}, and phase transitions that occur.  In particular, we are
interested in the heavy and light tailed degree distributions of
subnetworks. Concluding remarks and applications of the dynamic
rules investigated in this paper to the design of P2P networks are
provided in Section~\ref{sec-conc}.

\section{Model Parameters}\label{s1}
A list of the parameters and variables used in this paper,
along with their definitions, can be found in Table
\ref{tb1}. Throughout this paper, $q\in \{1,2,...,Q\}$ will represent
the type or class or category (all three terms are used
interchangeably) of a node in a network in which nodes can belong
to one of $Q$ different classes.

The dynamics for network evolution considered in this paper can be summarized as follows:\\
  (i) {\em Addition of Nodes:} At each time step, a new node is introduced into the
network, the new node can belong to any of the $Q$ different
classes or types indexed by an integer $q=1,2,...,Q$. The
probability of the new node being of type $q$ is assumed to be
$s_q$, where $\displaystyle\sum_q s_q=1$. \\
(ii) {\em Creation of Links:} The new node, inserted at time step $t$, then makes $m$ connections
by picking nodes preferentially using the well known linear kernel. Target nodes, however, can refuse requests for
connections, and hence, the new node performs the
following procedure $m$ times: It chooses a node preferentially;
thus the probability of choosing a node $i$ in the $q^{th}$ class is
$\displaystyle \frac{k(i,t;q)}{\sum_{j,p}k(j,t;p)}$ where $k(i,t;q)$ is the degree of the
$i^{th}$ node in class $q$, at time step $t$ (see Table~\ref{tb1}). The new node then sends a connection request
 to this selected candidate. The candidate node $i$
will accept the connection with probability $d_q$ depending on its
type $q$. If the connection is refused then, {\em the new node has
to repeat the process until it finds a node that accepts a new
connection}.  \\
(iii) {\em Deletion of Nodes:} At each time step, for each class $q=1,2,...,Q$
a randomly selected node of type $q$ and all its links are
deleted with probability $c_q$. Thus, the total deletion rate is $c=\sum_{i=1}^Q c_q <1$.\\
(iv) {\em Compensation for Lost Edges:} If a node looses a link
due to deletion of one of its neighbors, a node of type $q$ will
introduce $n_q$ new links following the same linear preferential procedure
outlined above  in step (ii).

 An important characteristic of this model is that
it is {\em local and private} in the sense that only a node
itself, and not the other members including the nodes trying to
connect to it, has any knowledge about its type.

{\em The parameters $s_q, d_q, c_q$, and $n_q$ thus represent the
heterogeneity} in the population, attraction/attachment, stability and
responsiveness dynamics that characterize the different categories of nodes:
\vspace*{-6pt}
\subsection{Nonuniform Attraction ($d_q$)}\vspace*{-4pt}
 Different
categories of nodes might have different degrees of attraction
(also known as fitness in \cite{d3}) that can influence the
structure of all the classes. That is, a class of nodes might be more
willing to accept the requests for new connections than the
others.

\vspace*{-6pt}
\subsection{Nonuniform Stability ($c_q$)} \vspace*{-4pt} The degree of stability of the nodes in different classes
might be different. In an ad-hoc network, where nodes can join and
leave the network, some classes of nodes might be inherently more
stable than others. Those classes, by the virtue of the fact that
they would stay longer periods of time in the network, will tend
to acquire larger fractions of the connections and usually tend to
become more heavy-tailed. The interaction of the classes of
"older" nodes and the class of "fresh" nodes is  fairly
interesting. The phases of the subnetworks can be tracked
separately for instance to determine the situations in which the
subnetwork of "old" nodes will acquire almost all the links of the
network.

\vspace*{-6pt}
\subsection{Nonuniform Responsiveness ($n_q$)} \vspace*{-4pt}The degree of responsiveness of different sets
of nodes to changes in the network might be different. We examine the effects of
heterogenous compensation magnitudes for the lost links. In a
compensation mechanism as introduced in \cite{us}, a node will react
to losing a neighbor by initiating a number of new connections to
\emph{compensate} for its lost links. The number of such
compensatory connections is an indication of the degree with which
the node responds to its changes. We show that the different degrees of responsiveness in
different classes will influence the structure of other classes
and the overall network.

\vspace*{-6pt}
\subsection{Nonuniform Population Size ($s_q$ and $c_q$)} \vspace*{-4pt}
As stated in
Table~\ref{tb1}, the number of nodes of type $q$ at time $t$ in
the network is given by $N(t;q)=(s_q-c_q)t$.  Hence, by varying
$s_q$ and $c_q$ the proportions of different classes of nodes can
be varied.  In the special case of two classes one can define a
majority and a minority class, and then study the effect of the
relative populations of the different classes on the PL exponent
of the degree distributions of each class. We derive in
Sections~\ref{s2} and \ref{s3}, the role that sizes of the
majority and minority classes play in determining the overall
network structure.

\begin{center}
\begin{table*}
\caption{Nomenclature}
\begin{tabular}{|c|c|c|}
  \hline
  Variable& Definition& Relation\\
  \hline
  \hline
  $Q\quad$ &  the number of different types of nodes &\\
  $q\in\{1,2,...,Q\}\quad$
  & denotes a particular class&\\
  $m\quad$ & number of links per newly inserted node &\\
  $s_q\quad$ & is the fraction of nodes of type $q$, per newly inserted& $\sum_q s_q=1$ \\
  $c\quad$& the fraction of nodes deleted per an inserted node&\\
  $c_q\quad$ & the fraction of nodes of type $q$ deleted per an inserted
  node &$\sum_q c_q=c$\\
  $d_q\quad$ & the probability that a node of type $q$ accepts a request
  for connection.& \\
  $m_q$& the probability of requesting a node of class $q$ for
  connection&$m_q=L(t;q)/L(t)$\\
  $\delta_q\quad$ & the probability of a new link to be finally connected to a
  node of class $q$& $\delta_q=\frac{m_q d_q}{\sum_{p=1}^Q (m_p d_p)}$ . \\
  $L(t;q)\quad$ & the sum of the degree of all nodes of type $q$ ($\sum_{i}k(i,t;q)$).& \\
  $L(t)\quad$ & the sum of the degree of all nodes in the network (twice the number of
  all links)&$L(t)=\sum_q L(t;q)$\\
  $k(i,t;q)\quad$ & the degree of a node $i$ of type $q$ at time step
  $t$.&\\
  $D(i,t;q)\quad$ & the probability that a node of type $q$ inserted at
  time $i$ is still in the network at time $t$.& $D(i,t;q)=(\frac{t}{i})^{-\frac{c_q}{s_q-c_q}}$\\
  $\gamma_q\quad$ & the power-law exponent of the scale-free degree
  distribution in class $q$ ($p_q(k)\propto k^{-\kappa}$).&\\
  \hline
\end{tabular}\label{tb1}
\end{table*}
\end{center}

\section{Nonuniform Attraction and Population}\label{s2}
We consider the case where the different classes are characterized
by different values of $d_q$ (acceptance probability) and $s_q$
(population size). We assume that there is no compensation, i.e.,
$n_1=n_2=\cdots=n_Q=0$, and that the deletions of nodes are made
uniformly over all the classes, i.e., $c_q=c*s_q$. Note that the
homogeneous case, where there is only one class, i.e., $|Q|=1$,
was solved in \cite{us}.  Let $i_{t;q}$ be the set of all nodes of
type $q$ that are present in the network at time $t$. When a new
link chooses a node $i\in i_{t;q}$ for connection, $i$ can accept
to attach to it with some probability $d_q$ and deny the
attachment with probability $1-d_q$. Once denied, the new link
will have to repeat the process to choose another target node for
connection.

We first provide the following reduction: the protocol for
selecting target nodes globally preferentially,  until a node is
found that accepts the connection, {\em is equivalent to} first
selecting a class $q$ with probability  $\delta_q$, and then
making a connection to the $i^{th}$ node in the class with
probability proportional to its degree as normalized with respect
to the sum of the degrees of all nodes {\em only in the $q^{th}$
class}, i.e., the probability that the $i^{th}$ node in class $q$
will receive an edge is $\displaystyle \frac{\delta_q
k(i,t;q)}{L(t;q)}$, where $L(t;q)$ is the sum of the degrees of
nodes in class $q$ (see Table~\ref{tb1}). In the equivalent
protocol,  the process of acceptance and denial is captured by the
parameter, $\delta_q=f(d_q,s_q)$, which is the steady state
probability of the new link being {\em finally connected} to a
node of type $q$; the relationship among $\delta_q, d_q,s_q$ is
derived later in this section. The modified protocol is derived to
make our analysis simpler.

Next we prove the equivalence of the modified protocol (where the
incoming node needs to select a class first, and hence, require
global knowledge) to the original protocol (where the incoming
node has no knowledge of the different class). Let (i)$A_q$ be the
event that a node of type $q$ is \emph{the end node} of a
successful link attempt; hence, $p(A_q)=\frac{L(t;q)}{L(t)}=m_q$,
(ii) $C_{i;q}$ be the event that a node $i$ of type $q$ is {\em
requested} for a connection. Then, $p(\hbox{a new link is
established}| C_{i;q})=d_q$, and
$p(C_{i;q})=\frac{k(i,t;q)}{L(t)}$. Also, $p(\hbox{a new link is
established })=\sum_{p=1}^Qm_pd_p$, and we are interested in
$p(C_{i;q}|\hbox{a new link is made})$. Using Bayes' rule, we get:
\begin{eqnarray} \label{eq:rate}
p(C_{i;q}|\hbox{ a new link is made})&\hspace*{-3cm} & \nonumber \\
&\hspace*{-3cm} =& \frac{p(\hbox{ a new link is established}| C_{i;q}) p(C_{i;q})}{p(\hbox{ a new link is established })}\nonumber\\
& \hspace*{-3cm} =& \frac{d_q k(i,t;q)}{L(t)(\sum_{p=1}^Qm_pd_p)}\nonumber\\
&\hspace*{-3cm} =& \frac{d_q m_q}{\sum_{p=1}^Qm_pd_p} \frac{k(i,t;q)}{L(t;q)}\nonumber\\
&\hspace*{-3cm} =& \delta_q \frac{k(i,t;q)}{L(t;q)}\ .
\end{eqnarray}
The expressions for $L(t;q)$ and $m_q$ are derived later in this section.

The continuous rate equation approach \cite{Dog3,us} can now be
employed to track $k(i,t;q)$, at a time $t\geq i$ in the equivalent protocol:
\begin{equation}\label{eqn11}
\frac{\partial k(i,t;q)}{\partial
t}=\frac{m\delta_qk(i,t;q)}{L(t;q)}-c \frac{k(i,t;q)}{N(t)},
\end{equation}
where {\bf (i)} the first term represents the fraction of new links that
the $i^{th}$ node of type $q$ gets due to the addition of $m$ new links at each step (see Eq.~(\ref{eq:rate})), and {\bf (ii)} the second term represents the fraction of edges lost due to the deletion of a randomly picked node at each time step. Note that, in general $N(t;q)=(s_q-c_q)t$, and since in the case of uniform deletion $c_q=c*s_q$, we get $N(t;q)=s_q(1-c)t$.
Similarly, a rate equation for  $L(t;q)$ can be written as:
\begin{eqnarray}\label{eq2}
 \frac{\partial L(t;q)}{\partial t}& \hspace*{-1.5cm}& \nonumber\\
&\hspace*{-1.5cm}= &m(\delta_q+s_q)-c
L(t;q)/N(t)-c\frac{L(t;q)}{L(t)}\frac{L(t)}{N(t)} \\
&\hspace*{-1.5cm}=&m(\delta_q+s_q)-L(t;q) \frac{2c}{(1-c)t},\nonumber
\end{eqnarray}
where {\bf (i)} the first term in Eq.~(\ref{eq2}) captures the following dynamic:
A new node brings $m$ links to the network. With probability $s_q$
this new node is of type $q$ and with probability $\delta_q$ one of
its ends will be connected to a node of type $q$; {\bf (ii)} The second term in Eq.~(\ref{eq2}) captures the following dynamic: When a node is deleted, it
might be of type $q$ with probability $N(t;q)/N(t)$, and the
class $q$ will lose an average of $L(t;q)/N(q,t)$ links, resulting
in an average contribution to $L(t;q)$ of $-c(N(t;q)/N(t)) \times
L(t;q)/N(t;q)=-cL(t;q)/N(t)$; and {\bf (iii)} the third
term in Eq.~(\ref{eq2}) corresponds to the links that nodes of class $q$ lose due to
their neighbors being deleted: When a node is deleted, an average number of
$L(t)/N(t)$ edges are deleted; now the fraction of these edges that are connected to
nodes of type $q$ is $L(t;q)/L(t)$.

$L(t;q)$ is then found to be:
\begin{equation} \label{eq:ltq}
L(t;q)=m(\delta_q+s_q)\frac{1-c}{1+c}t\ .
\end{equation}
Inserting it back into (\ref{eqn11}) we get,
\begin{eqnarray*}
\frac{\partial k(i,t;q)}{\partial t}
&=&\frac{\delta_q k(i,t;q)(1+c)}{(\delta_q+s_q)(1-c)}-c
\frac{k(i,t;q)}{(1-c)t}\\
&=&\frac{(\delta_q-c s_q)k(i,t;q)}{(\delta_q+s_q)(1-c)t},
\end{eqnarray*}
which implies $k(i,t;q)=m(t/i)^{\beta_q}$, for:
\[
\beta_q=(\delta_q-cs_q)/[(1-c)(\delta_q+s_q)].
\]
Next, using the relationship $(\gamma-1)\beta=1/(1-c)$ developed
in \cite{us} we get
\begin{equation}\label{eq10}
\gamma_q-1=\frac{(\delta_q+s_q)}{\delta_q-cs_q}.
\end{equation}

At this point, we can make several observations about the coexistence of different {\em phases} for the different classes of nodes and the achievability of different exponents for the different classes by varying the attraction rates.

\subsection{Tuning exponents $\gamma_q$'s and attraction probabilities $d_q$'s}
Recall that in Eq.~(\ref{eq:rate}) we derived the following relationship:
\begin{equation} \label{eq:delta}
\delta_q = \frac{m_q d_q}{\sum_{p=1}^Q m_p d_p},
\end{equation}
where $m_q=L(t;q)/L(t)$. Note that $L(t)=\sum_q L(t;q)$ $=mt(1-c)/(1+c)$, and hence using the result of
Eq.~(\ref{eq:ltq}), we get
$m_q=L(t;q)/L(t)=(\delta_q+s_q)$, which gives:
\begin{equation}\label{eqn2}
\delta_q= \frac{d_q(\delta_q+s_q)}{\sum_{q'} (\delta_{q'}+
s_{q'})d_{q'}}
\end{equation}
Given $d_q,s_q,$ for $q=1,2,...,Q$, (\ref{eqn2}) defines $Q$
equations which can be solved for the $Q$ unknowns $\delta_q$. Thus, without loss of generality, we  may assume that $\delta_q$'s are known and unique for any given set of $d_q$'s and $s_q$'s. Hence, we can uniquely find $\gamma_q$'s, for a fixed set of $d_q$'s, $s_q$'s and $c$.

Conversely, for a fixed set of $\delta_q$'s and $s_q$'s, the set of equations in (\ref{eqn2}) is linear in
$d_q$ (multiplying by the denumerator of the right hand side) .
Arbitrarily normalizing $d_q$'s to define a new set of variables
$d'_q=d_q/d_1$, $d'_q$'s can be found by solving the following
linear system of $Q-1$ equations:
\[
\sum_{q'=1}^Q d'_q(\delta_{q'}+s_{q'})(\delta'_q-I_{q;q'})=0
\]
for $q=2,3,...,Q$ and setting $d'_1=1$. The above system of
equations is easily shown to be non-singular if and only if
$\delta_q+s_q>0$ for all $q=1,2,...,Q$. In fact the determinant of
the system of equations is easily shown to be $\prod_{q=2}^Q
(\delta_q+s_q)$. {\em This shows the existence of attraction probabilities $d_q$ such
that any set of power-law exponents as constrained by Eqs.~(\ref{eq10}) and (\ref{eq2}) can be
achieved.}

\subsection{Co-dependencies of the feasible PL exponents for different classes}
As discussed above, by sweeping over the parameters $s_q$'s and $d_q$'s one can
effect a continuum of power-law exponents $\gamma_q$'s, as determined by Eqs.~(\ref{eq10}) and (\ref{eqn2}). {\em The set
of possible $\gamma_q$'s} are however coupled through the
requirement that $\sum_q \delta_q=\sum_q s_q =1$. As an example, one can define two  average quantities involving the power-law exponents:
\begin{equation}\label{eq12}
E_{node}(\gamma)=\sum_{q=1}^Q
s_q \gamma_Q \hbox{ and } E_{link}(\gamma)=\sum_{q=1}^Q
2^{-1}(\delta_q+s_q) \gamma_q\ \ ,
\end{equation}
where $E_{node}$ captures the fact that the averaging is taken
over randomly chosen nodes (i.e., probability that the class of a
randomly chosen node has exponent $\gamma_q$ is $s_q$), and
$E_{link}$ captures the fact that the averaging is done over
random endpoints of a randomly chosen edge, i.e., the probability
that a random endpoint of a randomly chosen edge belongs to the
$q^{th}$ class is $(\delta_q+s_q)/2$.  Now it follows from
Eq.~(\ref{eq10}) that
\begin{equation}\label{eq12}
E_{link}((\gamma-1)^{-1})=\sum_{q=1}^Q
2^{-1}(\delta_q+s_q)(\gamma_q-1)^{-1}=(1-c)/2\ .
\end{equation}
For
the special case of no deletion ($c=0$) it follows from (\ref{eq10}) that:
\[
E_{node}((\gamma-2)^{-1})=\sum_{q=1}^Q (\gamma_q-2)^{-1} s_q=1
\]
The convexity of the function $f(x)=1/x$ :
\[
E_{nodes}\{(\gamma-2)^{-1}\}\geq (E_{nodes}\{\gamma\}-2)^{-1}=1
\]
or $E_{nodes}\{\gamma\}\geq 3$. In other words, on average the
communities of the nodes have power-law exponents greater than 3.
On the other hand, this also implies that {\em even for the simple
preferential attachment,  heterogeneity can lead to an overall
degree distribution with exponent $<3$.} 

\begin{figure}
\begin{center}
\includegraphics[width=3.0in,height=2.0in]{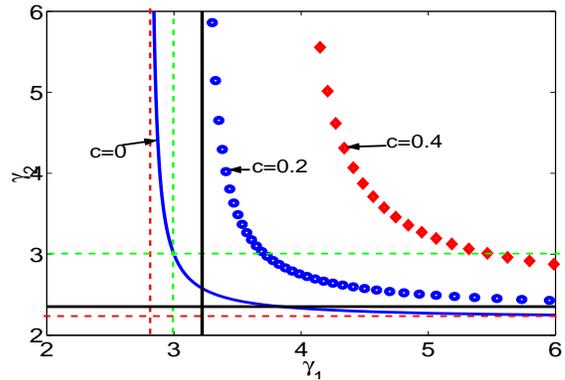}
\caption{ Possible power-law exponents in a network with two
classes of nodes. $\gamma_1$ is the power-law exponent of the
degree distribution of the first class that consists of $80\%$ of
all nodes, while $\gamma_2$ correspond to the second class
consisting of $20\%$ of all nodes. The possible power-law
exponents for three deletion rates $c=0,0.2,0.4$. The asymptotes
for $c=0$ and $c=0.2$ are also depicted.}\label{fig-gamma12}
\end{center}
\end{figure}

 When $Q=2$, the two exponents
$\gamma_1(d_1,d_2),\gamma_2(d_1,d_2)$ can be explicitly found as
functions of $d_1,d_2$. By eliminating $d_1,d_2$, the set of
possible power-law exponent pairs $(\gamma_1,\gamma_2)$ can be
derived. An example of this is depicted in Fig. \ref{fig-gamma12}
for $s_1=0.8,s_2=0.2$ for different values of $c$. To interpret
the asymptotes, first note that from (\ref{eq10}), stable network
operation is
 only possible when $\delta_q>cs_q$, otherwise the class $q$
 will become extinct (i.e., will loose all its links). Take class $1$ for instance.
 Since $\delta_2>cs_2$, then $\delta_1=1-\delta_2 < 1-c
 s_2=1-c+cs_1$, which results in: $\gamma_1 \geq 1+
(1-c+cs_1+s_1)/(1-cs_2-cs_1)=2+s_1(1+c)/(1-c)$. The same bound can
be found for
 class $2$ as well. Some of these asymptotes are also depicted in
 Fig. \ref{fig-gamma12}.

\subsection{Role of deletion rate $c$}
It was shown in \cite{us} that when $|Q|=1$, i.e., there is only
one class and the network is homogeneous, then $\gamma$ is always
greater than 3 for any deletion rate $c>0$. Moreover, even for
small values of $c$ the exponent becomes quite large, and the
network shows none of the characteristics associated with
heavy-tailed PL degree distributions. In the case of heterogeneous
networks, however, one can have a class with a true heavy-tailed
degree distribution (as illustrated in  Figs. \ref{fig-gamma12}
and ), and the overall network will thus exhibit a heavy-tailed
distribution, even for non-zero deletion rate $c$. However, as
$c\rightarrow 1$, we get (see Eqn. \ref{eq12}): $\sum_q
(\delta_q+s_q)(\gamma_q-1)^{-1}=0$. Thus, in the limit of heavy
node deletion, the degree distribution of any class $q$ with a
finite size ($s_q>0$) becomes exponential (i.e.,
$\gamma_q\rightarrow \infty$ for all $q=1,...,Q$).

In the next section, a compensatory mechanism is examined,
which will result in classes with prescribed heavy or light tailed
degree distributions, even in the limit of very high deletion rates.
Moreover, while the deletion process in this section was independent
of the class of nodes (a node was randomly chosen for deletion), in
practice, different classes of nodes will have different stability
characteristics. This important generalization is made in the
following section.

\section{Heterogenous Stability and Responsiveness}\label{s3}
Many ad-hoc networks are characterized by the fact that the time
scale within which their size grows, is much larger than the time
scale within which the nodes join and leave the network. In such
networks (certainly including P2P file sharing systems) one has the
case that $c\approx 1$. The scale free properties of growing networks that are subject to
permanent deletion of their nodes are studied in \cite{Dog3,us}.
In \cite{us}, the heavy-tailed structure of the resulting
scale-free networks were shown to immediately disappear in the
presence of node deletion. Results in the previous section also
showed that no heavy tailed structure can exist in the limit of
high deletion rates even in a heterogenous network.

 A universal compensatory protocol has been introduced by the authors
in \cite{us}, which ensures that the heavy tail of the degree
distribution of the emerging network is conserved even in the
limit of very high node departure rates. This section will
investigate the behavior of different classes of a network of
multiple types in the presence of {\bf (i)} \emph{heterogenous}
node deletion or equivalently, {\em heterogeneous stability}
factors, and {\bf (ii)} {\em heterogeneous responsiveness}, i.e.,
the rate at which different classes of nodes compensate for their
lost or dead links is class dependent. A class dependent
compensatory mechanism is a generalization of the universal
compensation scheme in \cite{us}, and it plays a crucial role in
restoring the heavy-tailed structure of some or all of the classes
in the network.

 The dynamical model introduced in Section \ref{s1} allows for non-uniform deletion of nodes
 and compensation of links. For simplicity we would assume
 uniform attraction, i.e., $d_1=d_2=...=d_Q=1$, that is, all nodes accept all requests for connections. We let the other
 parameters $s_q,n_q,c_q$ be arbitrary. The goal is to
 characterize the emerging scale free state of each of the $Q$
 sub-networks as a function of these dynamical parameters.

 \subsection{Rate Equation Formulation}
A mean-field  rate of change of $k(i,t;q)$ can be written as:

\begin{eqnarray}\label{l2}
     &&\frac{\partial k(i,t;q)}{\partial
    t}= \frac{mk(i,t;q)}{L(t)}-c(1-n_q)\frac{k(i,t;q)}{N(t)}\nonumber\\
    &+&\frac{k(i,t;q)}{L(t)} \left(\sum_{q'} n_{q'} \frac{L(t;q')}{L(t)}\right)\left( \sum_{q''}
    c_{q''}\frac{L(t;q'')}{N(t;q'')}\right)
     \nonumber\\
\end{eqnarray}

 where, (i) the first term comes from
the contribution of the $m$ new links inserted at time step $t$
when $L(t)$ is the sum of the degree of all nodes of type $q$
(remember that we assume uniform preferential attachment). (ii)
The second term captures the effect of the random deletion of one
of the neighbors of $i$ which is compensated with $n_q$
preferentially targeted new links (hence the $1-n_q$ multiplier).
(iii) The third term is due to the attraction of the compensatory
links from other nodes rather than $i$. It assumes that an average
of $\left( \sum_{q''} c_{q''}\frac{L(t;q'')}{N(t;q'')}\right)$
links are deleted, of which a fraction of $\frac{L(t;q')}{L(t)}$
belongs to a particular class $q'$ and will be compensated by
$n_q'$ preferentially targeted new links.

To find $L(t;q)$, its rate of change can be tracked as below:
\begin{eqnarray}\label{l22}
\frac{\partial L(t;q)}{\partial
t}&=&\nonumber\\
&=& ms_q+m\frac{L(t;q)}{L(t)}-c_q(1-n_q)\frac{L(t;q)}{N(t;q)}\nonumber\\
&-&\frac{L(t;q)}{L(t)}\left(\sum_{q'} c_{q'}
\frac{L(t;q')}{N(t;q')}\right)\nonumber\\
&\times& \left(1-\sum_{q''}n_{q''}
\frac{L(t;q'')}{L(t)}\right)\nonumber
\end{eqnarray}

where (i) the first terms corresponds to the new links added to
the class $q$ if the new node happens to be of type $q$ (this
occurs with probability $s_q$), (ii) the second  terms is due to
the end of new links being connected to a node of type $q$. (iii)
The third term comes from the deletion of an average of
$L(t;q)/N(t;q)$ links if a node of type $q$ is deleted, which is
compensated by $n_q$ new links per lost link. (iv) The forth term
is similar to the third term in (\ref{l22}). In the steady state
one has $L(t;q)\approx B_q t, N(t;q)=(s_q-c_q)t, N(t)=(1-c)t$. The
$Q$ unknowns $B_q$ can thus be found through the following set of
$Q$ equations (one for each $q=1,2,...,Q$):
\begin{eqnarray}\label{bqs}
B_q&=&\\
&=& m \left(s_q+\frac{B_q}{\sum_{q'}
B_{q'}}\right)\nonumber\\
&-&\frac{c_q}{s_q-c_q}(1-n_q)B_q\nonumber\\
&-&\frac{B_q}{\sum_{q'} B_{q'}} \left(1-\sum_{q''}n_{q''}
\frac{B_{q''}}{\sum_{q'} B_{q'}}\right)\left(\sum_{q''}
\frac{c_{q''}}{s_{q''}-c_{q''}} B_{q''}\right)\nonumber
\end{eqnarray}

Finding $B_q$'s and inserting back into (\ref{l2}), $k(i,t;q)$ is
found to be:
\begin{eqnarray}\label{kb}
k(i,t;q)&=&m(t/i)^{\beta_q}
\end{eqnarray}
for $\beta_q$ a function of $B_q,c_q,s_q,n_q$'s as below:

\begin{equation}\label{eqn:beta}
\beta_q=\frac{m}{B}-\frac{c(1-n_q)}{s_q-c_q}+\frac{\sum_{q'}n_{q'}B_{q'}}{B}\times
\sum_{q''}\frac{c_{q''}B_{q''}}{s_q-c_q}
\end{equation}
where $B=\sum_{q} B_q$.

Now define $D(i,t;q)$ the probability that a node of type $q$,
inserted at time $i$ is still in the network at time $t$.
$D(i,t;q)$ can be found as follows:
\begin{eqnarray*}
\frac{\partial D(i,t;q)}{\partial t}&=&-c_q
D(i,t;q)/N(t;q)\\&=&-c_q D(i,t;q)/[(s_q-c_q)t]
\end{eqnarray*}
resulting in, $D(i,t;q)=(t/i)^{-c_q/(s_q-c_q)}$.

Finally, $P_q(k)$, the degree distribution of the nodes of type
$q$ can be found as follows:
\begin{eqnarray}\label{deg}
    P_q(k)& = & \frac{\hbox{No. of nodes of type $q$ with degree $=\ k$}}{\hbox{Total number of nodes of type $q$}}
    \nonumber \\
    &=& \frac{1}{N(t;q)}\sum_{i:k(i,t;q)=k}
    N(t;q)D(i,t;q)\nonumber \\
    &=&\displaystyle D(i,t;q)\left\vert\frac{\partial k(i,t;q)}{\partial
    i}\right\vert_{i:k(i,t;q)=k}^{-1}
\end{eqnarray}

Solving for $i$, such that $k(i,t;q)=k$ from (\ref{kb}), and
inserting back into (\ref{deg}) we arrive at:
\begin{eqnarray*}
P_q(k)\propto k^{-c_q/[(s_q-c_q)\beta_q]-1/\beta_q-1}\propto
k^{-\gamma_q}
\end{eqnarray*}
from which the power-law exponents are found to be:
\begin{eqnarray}\label{gammaqs}
\gamma_q
 &=& 1+\frac{s_q}{\beta_q(s_q-c_q)}\nonumber\\
\end{eqnarray}

\begin{figure}
\begin{center}
\includegraphics[width=3.0in,height=2.0in]{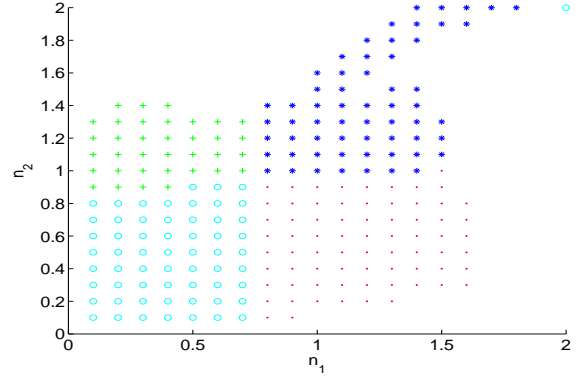}
\caption{Different regions of the power-law exponents for an
example consisting of two classes of nodes with $s_1=0.8, s_2=0.2,
c_1=0.6, c_2=0.15$ for various compensation magnitudes
$0<n_1,n_2<2$. (i) The circles indicate the region in which class
one is light tailed while class two is heavy tailed. (ii) Squares
indicate the phase where both classes are heavy tailed (iii) both
classes are light tailed where indicated by diamonds and (iv)
asterixes indicate that the first class is heavy tailed while the
second class is light tailed.}\label{fig-phases_new}
\end{center}
\end{figure}

\begin{figure}
\begin{center}
\includegraphics[width=3.5in,height=2.5in]{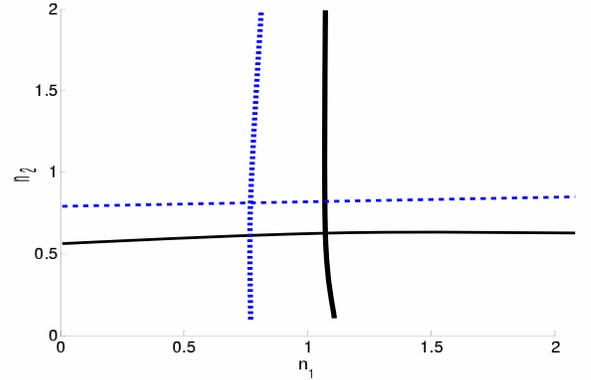}
\caption{Phase diagram of a mixture of two classes of nodes. The
entry rate of the two classes are the same $s_1=s_2=0.5$. The axes
correspond to $n_1,n_2$. The curves represent the points in
$(n_1,n_2)$ for which either $\gamma_1=3$ or $\gamma_2=3$. The
dashed curves are for the case of $c_1=c_2=0.3$ and are hence
symmetric. As the deletion rates becomes un-balances, the phases
curves also shift. The solid curves correspond to the deletion
rates $c_1=0.4,c_2=0.3$. It therefore takes more compensation for
the first class to retain its heavy tail. Now the second class in
its turn will acquire some of these excess compensations. Thus,
overall the class two will need less compensation to stay heavy
tailed. The net result is a shift of the phase diagram to the
south-east. }\label{fig-phases_2}
\end{center}
\end{figure}

Finding the set of $\gamma_q$'s requires the tedious task of
solving the system of equations (\ref{bqs}) to find $B_q$'s and
then plugging the $B_q$'s back into (\ref{eqn:beta}) to get the
$\beta_q$'s and finally finding $\gamma_q$'s through
(\ref{gammaqs}). Such procedures are carried out for a number of
examples in the next section.

\section{Phase Diagrams and Topological Observations}\label{sec-phases}

\subsection{Phase Diagrams}
There are four particularly interesting emergent phases for a
subclass of nodes of a network:
\begin{itemize} \item The Light-Tailed phase, in which the
network emerges into a quasi-equilibrium state in which the
average degree of the nodes is bounded and non-zero, and the
variance of the degree distribution is also bounded. \item The
Heavy-Tailed phase, corresponding to a subnetwork with finite
average degree but diverging variance of the degree distribution.
Such networks possess many attractive properties (at least in
their equilibrium form) like constant diameter or zero percolation
probability \cite{newman}. \item Extinction phase, when the
average degree of the subnetwork goes to zero and it looses most
of its links. \item Unstable or divergent phase, in which the
average degree of the subnetwork diverges and the whole network
loses its stability.
\end{itemize}

Previous sections suggest the procedure for calculating the
emergent power-law exponent of all the network classes as a
function of the model parameter, $n_q,c_q,s_q$. This in turn can
determine the state of all subnetworks of the network. An example
of such procedure  is carried out for two classes of nodes with
$s_1=0.8, s_2=0.2, c_1=0.6, c_2=0.15$ for various compensation
magnitudes $0<n_1,n_2<2$. Only the heavy-tailed and light-tailed
phases are marked and thus there are a total of 4 possible phases
for the whole network depicted in Fig. \ref{fig-phases_new}.

The contour in the space of parameters for which $\gamma_q=3$ is
important because it marks the transition of the phase of the
network from a stable light-tailed state into a stable
heavy-tailed one. An example is depicted in Fig.
\ref{fig-phases_2}, in the scope of parameters $n_1,n_2$ for fixed
$c_1,c_2,s_1,s_2$.

To our special practical interest is the case of quasi-uniscale
networks, the networks in which most of the nodes belong to a
single category and  most deletions and insertions happen in the
nodes of this category. Call this category, $q=1$. We then define
a \emph{quasi-uniscale} network as one in which $c_1\approx c \gg
c_q, s_1 \approx 1 \gg  s_q, s_1 - c_1 \gg s_q-c_q, q\neq 1$.

Class $1$ will be called the majority class and the rest of the
classes are called minority classes. It then follows that
$L(t,1)\approx L(t)$ and $L(t;q)\ll L(t)$ for all $q\neq 1$ which
reduces (\ref{bqs}) to: $L(t;q) \approx \delta_{q,1}
2m(1-c)t/(1+c-2n_1 c)$, where $\delta_{q;q'}$ is the Kronchker's
delta function. Inserting in (\ref{eqn:beta}), the $\beta_q$'s are
found to be:
\[
\beta_q \approx \frac{(1-c+2c(n_q-n_1+n_1\delta_{q,1}))}{2(1-c)}
 \]

from which:
 \begin{eqnarray*}
 \gamma_1\approx 1+\frac{2}{1-c+2cn_1}\\
 \gamma_q \approx 1+\frac{2 r_q }{1-c+2c(n_q-n_1)},
 \end{eqnarray*}
 where $r_q=\frac{(s_q-c_q)}{s_q(1-c)}$, called the robustness factor of the class $q$ is the fraction of nodes of
 type $q$ that are in the network over the fraction of nodes that
 enter per newly inserted node. In the limit of very active
quasi-uniscale networks ($c\rightarrow 1$), one gets : $\gamma_1
\approx 1+1/n_1$
 while $\gamma_q \approx 1+\frac{r_q}{n_q-n_1}$ for $q\neq 1$.

\begin{figure}
\begin{center}
\includegraphics[width=3.5in,height=3.0in]{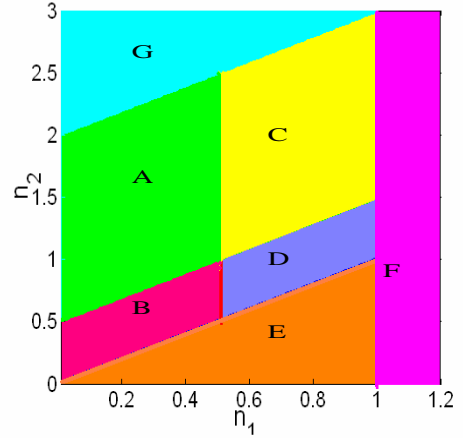}
\caption{ Different phase regions in the space of parameters
$(n_1,n_2)$, where $n_1$ is the compensation rate for the majority class
while $n_2$ is the corresponding parameter for the minority class.
The phases are as follows:
(A) The majority is light-tailed while the minority is
heavy-tailed. (B) Both the majority and minority are light-tailed
(C) Both are heavy-tailed (D) Minority is light-tailed while
majority is heavy-tailed (E) The minority extinction (F) Majority
becomes unstable (infinite mean) (G) Minority becomes unstable
(infinite mean)}\label{fig-phases-mm}
\end{center}
\end{figure}

An example of the phases for this limiting case is depicted in
Fig. \ref{fig-phases-mm}.

\begin{figure}
\begin{center}
\includegraphics[width=4.0in,height=3in]{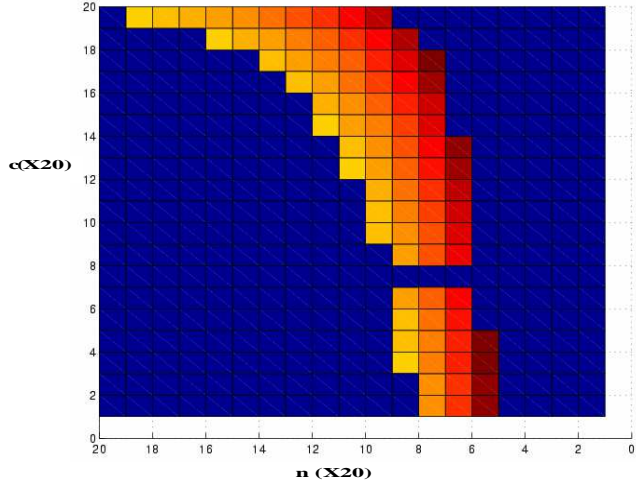}
\caption{Stable heavy-tailed minority class: The majority class
with $s_1=0.75,c_1=0.8$ is not compensating ($n_1=0$). The region
in the space of the $n_2,c_2$ where the minority class is stable
and heavy tailed (i.e., $2<\gamma_1<3$) is highlighted. The part
to the left of the region (for higher compensation rates)
indicates $\gamma_1<2$ while the region to the rights indicates
$\gamma_1\geq 3$.}\label{3D}
\end{center}
\end{figure}

Another interesting limit is where only one class compensates for
its lost links and the rest of the classes are irresponsive.
Consider the case of a mixture of two classes where the first
class with $c_1=0.8, s_1=0.75$ does not compensate ($n_1=0$). The
phases regions  of the second class with $s_2=0.25$ are depicted
in Fig. \ref{3D} as a function of its deletion and compensation
rates ($c_2,n_2$).

\subsection{Simulations}

\begin{figure}
\begin{center}
\includegraphics[width=3.0in,height=3.0in]{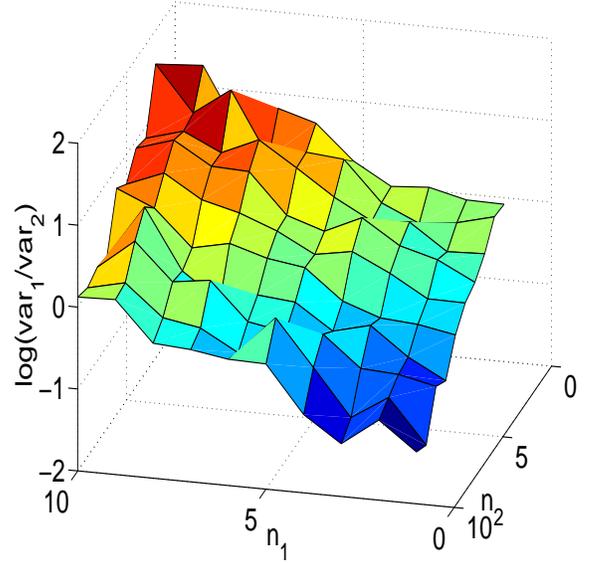}
\caption{The ratio of the variances of two classes with $c=0.5$,
$s_1=s_2=0.5$ as a function of the compensation magnitudes
$n_1,n_2$. The variance of the degree distribution increases as
the compensation magnitude of the corresponding class
increases.}\label{ratioc1c2}
\end{center}
\end{figure}

For a network of finite size, the variance of the degree
distribution is an indication of how heavy-tailed the degree
distribution is. Plotting the ratio of the variances at different
classes in the space of dynamical parameters will allow us to
compare the relative state of different network classes. We have
simulated a network of 5000 nodes with two categories of nodes
,$Q=2$, with uniform deletion of magnitude $c=0.5$ and
$s_1=s_2=0.5$ for various compensation magnitudes $n_1,n_2$. We
have then plotted the ratio of the variances at each of the
classes for each value of $(n_1,n_2)$ in Fig. \ref{ratioc1c2}.

\begin{figure}
\begin{center}
\includegraphics[width=3in,height=2.5in]{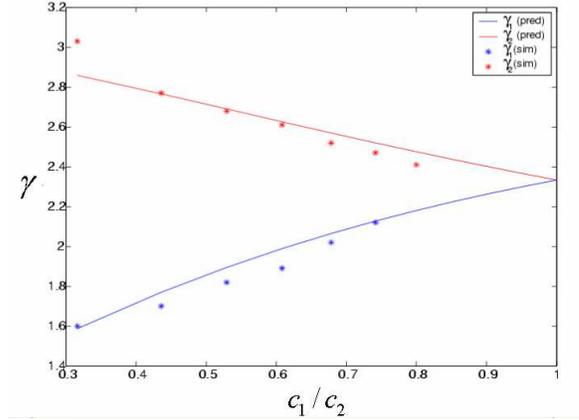}
\caption{The power-law exponent of a network of two classes with
$10000$ nodes. The insertion rates are $s_1=s_2=0.5$ and the
overall deletion rate is a constant $c=0.5$. The deletion rate of
the two classes are however different. For various ratios of the
deletion rates, the power-law exponent of these two classes are
found through simulation. Also depicted are the theoretical
predictions of the previous section.}\label{fig:sim}
\end{center}
\end{figure}

Simulations to obtain power-law exponents for two classes with
equal insertion rates and various deletion rates are depicted in
Fig. \ref{fig:sim} and verified against the analytical
expectations.

\subsection{Topological Observations}
As suggested in the introduction, the class of nodes that are more
heavy tailed are expected to play more central roles in the
topology of the network. The light tailed classes on the other
hand would be pushed to the edges of the network serving as leaf
nodes. This is a very desirable property for many application
including P2P communication systems as discussed in more details
in the following section. In this section we try to quantify the
\emph{place} of a category of nodes in the network.

\begin{figure}
\begin{center}
\includegraphics[width=3.0in,height=2.5in]{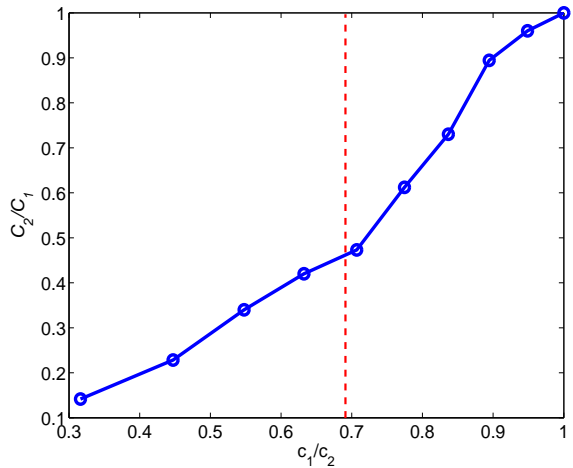}
\caption{The ratio of the capacities of the two node categories.
The network size is 10000, $s_1=s_2=1/2$ and $c=1/2$. The ratio of
the deletion rates for the two categories are varied from 0.3 to
1. The dashed line marks the transition of the first category from
a divergent phase into the phase with power-law exponent greater
than 2.}\label{fig:cap}
\end{center}
\end{figure}

The quantity we consider is the so called \emph{capacity} of each
subnetwork. For a node category $q$, the capacity ${\mathcal C}_q$
is defined as \emph{the total number of edges that have both their
ends in a node of type $q$, over twice the total degree of all the
nodes of type $q$}. When ${\mathcal C}_q=0$, the category $q$ has
all its edges to the outside of the category $q$. On the other
hand ${\mathcal C}_q=1$, all the links of the nodes in category
$q$ stay within the same category. One can thus assume that a
network with small capacity has a leaf role, while a large
capacity is an indication of a more compact topological structure.
Fig. \ref{fig:cap} depicts the capacities of the two categories of
a network of two categories as a function of the relative rate of
deletion $c_1/c_2$. The more stable subnetwork will have a  larger
capacity, which decreases as $c_1/c_2\rightarrow 1$.

\section{Concluding Remarks: Applications to P2P Networks}\label{sec-conc}
An important example of a complex, highly dynamic, and heterogeneous
network (which also partially motivated this research) is the less
structured or ad-hoc distributed systems with peer-to-peer (P2P)
content sharing networks as a prime example. While nodes in a P2P
computer network have heterogeneous resources and lifetimes, they
can be classified into a few meaningful classes based on hardware
or software characteristics. In particular, the nodes can be
categorized into two major classes \cite{limewire}: (i)Super-nodes
with virtually infinite bandwidth (e.g. office users) that run a
super-node software and (ii) the low bandwidth home users that run
the ordinary software. The fraction of super-nodes is extremely
small (around 1\% of the whole nodes) but super-nodes are much
more stable, with their life times ranging anywhere between 10 to
100 times of the home users.

 The integrity of such P2P networks
requires most communication paths to be provided by the
super-nodes; otherwise, the  traffic at the home users will soon
exceed their limits and the network structure will be fragmented.
This can be ensured only when the network core, that is the highly
connected nodes in the network, are mostly super-nodes (or nodes
with more capabilities). An ignorance of this fact has led to the
apparent break down of the Gnutella network, an early P2P file
sharing system in 2000 \cite{clip2}.

The dynamics of P2P networks is dominated by the rapid rate of the
members joining and leaving the network. More than $60\%$ of all
the nodes joining these networks will leave within the first hour,
while it takes around three months for the overall size of the
network to grow by $60\%$ \cite{clip2}. Ensuring the emergence of
a heavy tailed scale-free state in such ad-hoc environments is a
challenging task. As was shown in Section \ref{s3}, the same
compensatory mechanism developed in \cite{us} can ensure the
emergence of scale-free structures with heavy tails among more
stable groups, the groups anticipated to be composed of nodes with
high capacity, while the majority of the nodes will have a light
tailed degree distribution and are therefore exempted from the
search paths. {\em The results of this paper will serve as an
essential part of ad-hoc network formation protocols} that can
support efficient search \cite{perc,p2p,tcs}, robustness, and
allow highly dynamic operations.

In passive networks, in which the connections of a new node are
never redirected or modified once they are established, the
dynamical parameters ($s_q,n_q,d_q,c_q$) are usually fixed
constants. On the other hand, in most active networks (e.g; P2P
applications), some or all of these parameters can be locally
manipulated by the nodes. The manipulations of these parameters
can be considered as \emph{\bf active network design protocols} that
can modify these parameters and thus can engineer the power-law
exponents $\gamma_q$. Consider first the possibility of simulated
deletion of the nodes of a network. There are classes of networks
(e.g. P2P computer networks) in which a node can simulate its
departure (log-off) through a \emph{software} decision.  In such
networks, the stability of a class of nodes can be manipulated as
follows: At each time step, for any $q$, with probability $l_q$, a
randomly chosen node of type $q$ decides to leave the network (and
hence disconnects all its links) and quickly joins back (with $m$
preferentially targeted links). This would then modify the model
parameters as follows: $s_q\rightarrow (s_q+l_q)/(\sum_{q'}
l_{q'})$, $c_q\rightarrow (c_q+l_q)/(1+\sum_{q'} l_q)$. Also, the
compensation magnitude $n_q$ of any class can be manipulated
through the "software" running on that node. By varying $n_q$, the
class $q$ can tune its emergent power-law exponent $\gamma_q$.

\begin{figure}
\begin{center}
\includegraphics[width=3.5in,height=3.5in]{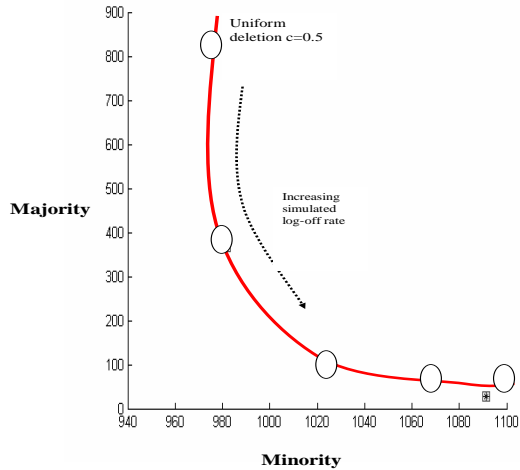}
\caption{{\bf Manipulating Network Structure By Altering Local Rules}: The change in the variance of the degree distribution in
a network consisting of a majority class ( 90\% of the nodes) and
a minority class. The network has size 10,000 and both classes
compensate with rate $n=1$. At the point indicated by "uniform
deletion", a uniform deletion with rate $c=0.5$ is in place.  The
path shows the change in the variance of the degree distribution
as the majority starts to simulate a log-off.  Without any
simulated log-off, the variance in both classes are high, but as the
simulated log-off rate increases, the variance of the majority
quickly decreases, while the minority remains heavy tailed.
}\label{fig-sim-logoff}
\end{center}
\end{figure}

Figure (\ref{fig-sim-logoff}) is an example of how the simulated
log-off can tune the heavy tailed structure of a class.

In conclusion, the structure of preferentially grown networks with
heterogeneous preference kernels has traditionally been
categorized with a single parameter, namely the power-law exponent
of the overall scale-free degree distribution of the network, if
such scale-free state emerges. We introduced a number of  local rules
that can tune the emergent scale-free states of different classes
and in particular can ensure heavy or light tailed distributions
within a particular class. These protocols dealt with four major
dynamical elements of the network: The linkage properties of the
network; The rate of departure of the nodes in the network;
the rate with which nodes of a certain class $q$ compensate for
the links they lose; and the rate at which nodes accept requests
for connections or links. Different phases of the emergent subclasses
under these local rules were characterized and the boundaries of
phase transitions were identified.

\end{document}